**Facile synthesis of fine-grained $CoFe_2O_4$ anchored on porous carbon for simultaneous removal of tetracycline and arsenite**

Yuwen Chen[a], Ke Zhu[a], Yizhe Huang[a], Xin Li[a], Zhikeng Zheng[a], Zhiwei Jiang[a], Di Hu[a], Ping Fang[c], Kai Yan[a, b, *]

[a] *Guangdong Provincial Key Laboratory of Environmental Pollution Control and Remediation Technology, School of Environmental Science and Engineering, Sun Yat-sen University, Guangzhou 510275, Guangdong, China*

[b] *Guangdong Laboratory for Lingnan Modern Agriculture, South China Agricultural University, Guangzhou 510642, China*

[c] *South China Institute of Environmental Science, Ministry of Ecology and Environment, Guangzhou 510655, Guangdong, China*

[*] *Corresponding author Email address: yank9@mail.sysu.edu.cn (K. Yan)*

**Abstract:** The coexistence of tetracycline (TC) and arsenite (As(III)) in livestock wastewater threatens public health, and the heterogeneous Fenton-like system is a practical approach for the simultaneous removal of TC and As(III). In this work, fine $CoFe_2O_4$ nanoparticles are facilely anchored on heretically porous carbon ($CoFe_2O_4$@PC) via a microwave-assisted calcination method and used for eliminating TC and As(III) via peroxymonosulfate (PMS) activation. The $CoFe_2O_4$@PC/PMS system exhibits excellent performance in the simultaneous removal of 96% of 10 mg/L TC and 86% of 50 μM As(III) within 30 min, whereas the porous carbon can boost the mass transfer efficiency and protect $CoFe_2O_4$ against Co and Fe ions leaching. The $CoFe_2O_4$@PC/PMS system maintains satisfactory activity and stability in different conditions, including a broad pH range, anions, and humic acid. Meanwhile, the $CoFe_2O_4$@PC with the strong magnetic property was readily recovered from the reaction solution. Moreover, the cooperation of radical (·OH, $·SO_4^-$ and $·O_2^-$) and nonradical ($^1O_2$) pathways was documented to play critical roles in enhancing the oxidation of TC and As(III), which effectively reduces the toxicity of degradation intermediates. This work offers a facile strategy for constructing spinel-based catalysts and opens a broad prospect for removing TC and As(III) in livestock wastewater.

## 1. Introduction

Tetracyclines (TC) as broad-spectrum antibiotics are widely used as bacterial inhibitors and growth promoters [1]. These are full of challenges to be removed entirely in water due to their biodegradability. Residual antibiotics can lead to increased microbial resistance and threaten human health [2,3]. On the other hand, heavy metals are also frequently found in feed additives and detectable in livestock wastewater [4]. The chemical forms of As affect its toxicity and mobility, with arsenite (As(III)) being more harmful and moveable compared to arsenate (As(V)). The oxidation of As(III) to As(V) is beneficial in reducing toxicity [4,5]. As(III) and TC are often concurrently detected in livestock wastewater [6]. In addition, TC and As(III) are commonly found together on rice fields because of inadequately treated livestock effluent [7,8]. Therefore, it is significant to develop approaches that can simultaneously remove TC and As(III).

Recently, considerable research efforts have been devoted to Fenton or Fenton-like oxidation in wastewater treatment. Particularly, peroxymonosulfate (PMS) is recognized as a promising Fenton-like oxidant. The key to activating PMS is the cleavage of the O-O bond (bond energy = 377 kJ $mol^{-1}$) to generate $\cdot SO_4^-$ radical, which has higher redox potential than ·OH, and it can react with $H_2O$ or $OH^-$ to produce more ROS including ·OH, $\cdot O_2^-$ and $^1O_2$. Physical approaches (e.g., heat, light, ultrasound) and chemical approaches (e.g., alkaline, metal ions) are limited to high energy costs and chemical expenses, respectively [9]. Compared with the above techniques, heterogeneous catalysts, especially transition metal oxides, are considered favorable PMS activators with competitive efficiency, energy consumption, and less metal leaching.

In comparison with monometallic oxides, bimetallic oxides with more efficient redox cycling of diverse metal species typically possess higher activity and stability [10,11]. Bimetallic spinel oxides, with a formula of $AB_2O_4$, have received extensive attention in PMS activation. Different metals (e.g., Co, Fe, Mn, Ni, Zn) occupied either tetrahedrally coordinated A sites or octahedrally coordinated B sites, forming a stable structure [12]. The electrical conductivity and PMS activation efficiency of spinel oxides can be significantly improved due to the inherent multivalency. Charge transfer between cations having multiple chemical states occurs via a leap process requiring low activation energy [13]. Various spinel oxides, including $MnFe_2O_4$ [14], $CoFe_2O_4$ [15], $CoMn_2O_4$ [16] and $NiCo_2O_4$ [17], proved effective in activating PMS [11,18,19]. Among these, $CoFe_2O_4$ has optimal catalytic activity because Co and Fe possess feasible redox potential, and strong Fe-Co interaction significantly inhibits divalent metal ion leaching [20]. Besides, spinel ferrites are considered ideal arsenic adsorbents due to their strong affinity for arsenic species [21], and large surface hydroxyl groups found on the surface of metal oxides also benefit arsenic adsorption and radical generation during PMS activation [22,23]. Therefore, it is promising to utilize $CoFe_2O_4$ as the prime catalyst to remove TC and As(III).

Direct calcination of $CoFe_2O_4$ nanoparticles often leads to aggregation, inhibiting active site exposure and decreasing catalytic activities. Given this, researchers have explored loading metal oxides on various support with large specific surface areas. In particular, carbon-derived supports, such as activated carbon, graphene oxide, and carbon nanotube, are widely studied owing to controllable morphology, high surface area, and excellent electric conductivity [24–28]. Compared with graphene and commercial activated carbon, biomass-derived porous carbon, obtained from the pyrolysis of agricultural wastes, can be considered sustainable and applicable catalyst support. For example, Dong et al. used lignin-derived biochar as support to prepare $CoFe_2O_4$@BC with the hydrothermal method, which can effectively activate peracetic acid to degrade sulfamethoxazole [29]. Compared to hydrothermal, direct solid-state synthesis can avoid the addition of excessive chemicals, and microwave-assisted calcination with high heating efficiency can drastically reduce reaction time [30] and is suitable for large-scale production.

Herein, the present work aims at the facile synthesis of $CoFe_2O_4$ with biomass-derived porous carbon as support ($CoFe_2O_4$@PC) by the combination of the benefits mentioned above, which was applied to activate PMS for the simultaneous oxidative removal of TC and As(III) from water bodies. The results of XRD and TEM characterizations proved that $CoFe_2O_4$ nanoparticles were uniformly anchored on porous carbon. In the degradation experiment, 96% of TC was degraded within 30 min and 86% of As(III) was oxidized into As(V) within 5 min confirmed by HPLC-ICP-MS analysis. Additionally, we investigated the effects of distinct factors on the performance of catalytic activity. Quenching tests and study of ESR, XPS, and LC-MS showed that the cooperation of radical and nonradical pathways effectively improved the oxidation efficiency of organic pollutants, and As(III) was oxidized into As(V) through the radical pathway. Toxicity evaluation showed that the catalytic system effectively reduced the toxicity of pollutants in water and mitigated the environmental impact. This work provides a promising route for the simultaneous removal of organic pollutants and heavy metals in water bodies.

## 2. Materials and methods

Details of the preparation and experiment are depicted in Supporting Information.

## 3. Results and discussions

### *3.1. Characterization of catalysts*

After synthesis, the morphology and structure of as-prepared PC, $CoFe_2O_4$, and $CoFe_2O_4$@PC were characterized via various techniques. The crystalline phases of PC, $CoFe_2O_4$, and $CoFe_2O_4$@PC were determined by X-ray diffractometer (XRD) (Fig. 1a). The diffraction spectrum for $CoFe_2O_4$ and $CoFe_2O_4$@PC showed six peaks at 18.3°,

30.1°, 35.4°, 43.0°, 57.0°, and 62.6°, corresponding to (111), (220), (311), (400), (511) and (440) of $CoFe_2O_4$ (JCPDS.No: 22-1086), respectively. The diffraction peaks of $CoFe_2O_4$@PC were sharp and intense, and no impurity peaks were seen, indicating that the high crystallinity of spinel oxides (Fig. 1b) was maintained during microwave calcination. Besides, the pattern for PC had only two broad peaks, indicating the presence of fully amorphous carbon. The results showed a successful synthesis of $CoFe_2O_4$@PC with facile and efficient microwave-assisted calcination.

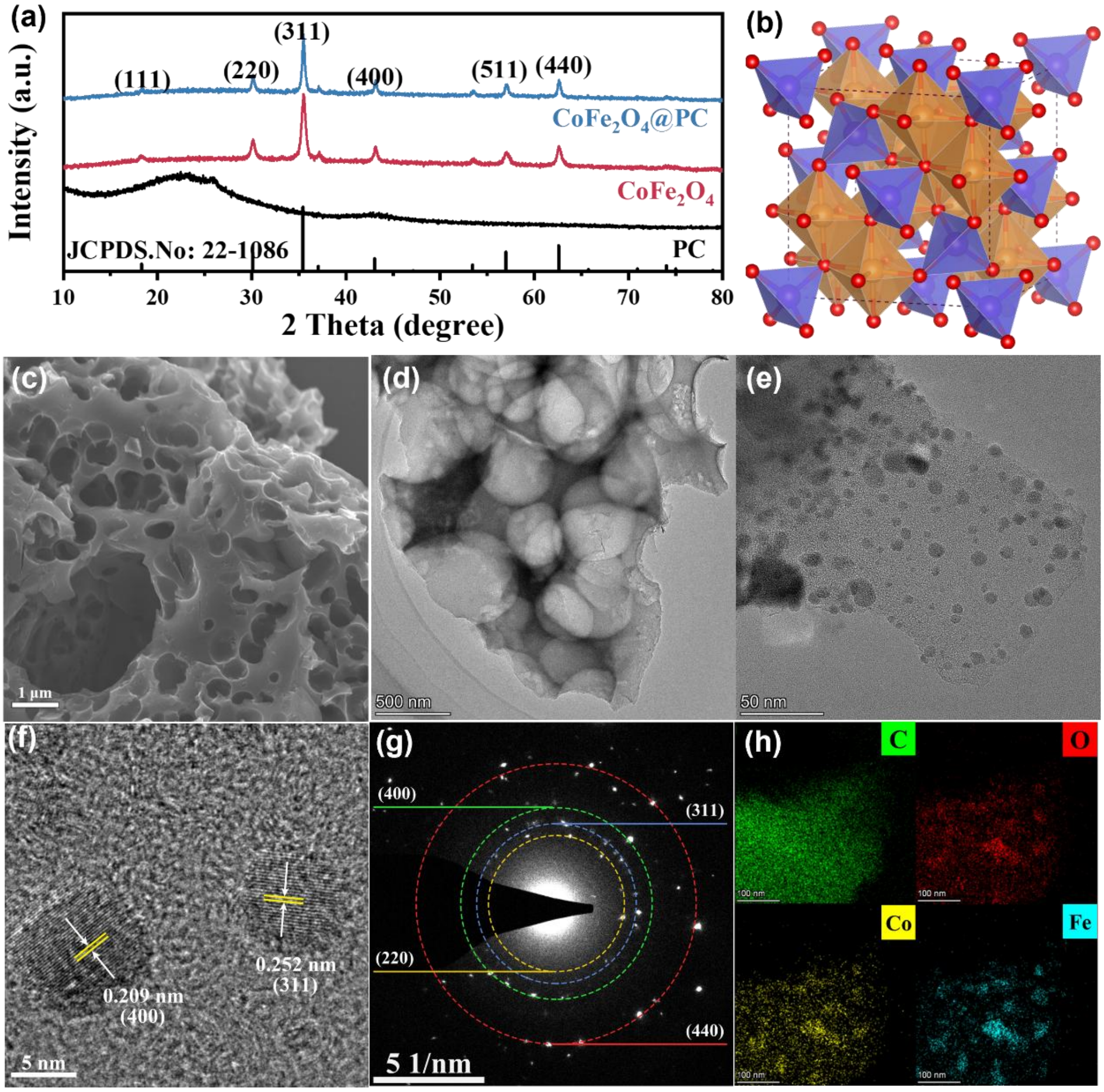

**Fig. 1.** Characterization of prepared catalysts. XRD pattern of PC, $CoFe_2O_4$, and $CoFe_2O_4$@PC (a); Crystal structure of $CoFe_2O_4$ (b); SEM image (c) and TEM image

(d) of PC; TEM image (e), HRTEM image (f), SAED pattern (g) and elemental mapping (h) of $CoFe_2O_4$@PC.

Scanning electron microscopy (SEM) and transmission electron microscopy (TEM) were then adopted to determine the morphology and loading of $CoFe_2O_4$. As shown in Fig. 1c-d, the SEM and TEM image of PC revealed that the stacks of thin-layer carbon were covered with porous structures with pore sizes ranging from 0.1-3 μm, which provides a larger specific surface area for spinel oxides loading and more active sites for pollutant removal. In more detail, the enlarged TEM image in Fig. 1e of $CoFe_2O_4$@PC clearly showed that the $CoFe_2O_4$ nanoparticles were finely anchored on the carbon layer, which implied a strong interaction between $CoFe_2O_4$ and carbon. Compared to the SEM image of $CoFe_2O_4$ in Fig. S1, $CoFe_2O_4$ anchored on PC exhibited smaller particle size, and no evident accumulation was found, which proved that PC could effectively prevent the collection of spinel oxides, thus providing more active sites for PMS activation. High-resolution transmission electron microscopy (HRTEM) was employed to further confirm the catalyst structure. Fig. 1f showed the lattice fringes spacing of 0.209 nm and 0.252 nm for the (400) and the (311) lattice planes of $CoFe_2O_4$. SAED pattern (Fig. 1g) showed that nanoparticles loaded on PC present the expected lattice spacings of 0.296 nm, 0.253 nm, 0.209 nm, and 0.148 nm for the (220), (311), (400), and (440) planes of $CoFe_2O_4$, respectively, coinciding with the observation from XRD results. Moreover, Elemental mapping in Fig. 2h demonstrated that $CoFe_2O_4$ was uniformly anchored on PC with high dispersion, further confirming the crucial role of PC in preventing agglomeration as support. Above all, $CoFe_2O_4$@PC with higher dispersion and more exposed reactive sites was successfully prepared with the microwave-assisted calcination method.

### 3.2. Catalytic performance

#### 3.2.1. TC removal

The catalytic performance of the as-prepared $CoFe_2O_4$@PC was first investigated with TC as the target pollutant. In the presence of 0.8 mM PMS and after 30 min of reaction (Fig. 2a), the removal efficiency of TC by PMS alone, $Co_3O_4$, $CoFe_2O_4$, and $CoFe_2O_4$@PC system attained 18.2%, 57.4%, 72.1%, and 96.5%, respectively. By fitting the pseudo-first-order kinetic model, corresponding TC degradation rate constants of PMS alone, $Co_3O_4$, $CoFe_2O_4$, and $CoFe_2O_4$@PC system were calculated to be 0.0067, 0.0280, 0.0396, and 0.0943 $min^{-1}$, respectively (Fig. 2b). It showed that the reaction rate of $CoFe_2O_4$@PC was 3.3, 2.4, and 14.1 times higher than that of $Co_3O_4$, $CoFe_2O_4$, and PMS alone, respectively. Compared to $Co_3O_4$, significant improvement in the $CoFe_2O_4$ system implies that the synergetic effect of Co-Fe in $CoFe_2O_4$ plays a crucial role in efficiently accelerating PMS activation. Furthermore, the enhanced activity in the $CoFe_2O_4$@PC system revealed the contribution of PC support for the following reasons: (1) PC effectively prevented agglomeration, exposing more reactive sites; (2) the porous structure of carbon reduced mass transfer resistance. Notably, about 20% of TC could be degraded in the PMS alone system, which may be due to the self-decomposition of PMS. Due to pH decreasing, metal ions could be leaching. A certain amount of $Co^{2+}$ was added to activate PMS, resembling the leached $Co^{2+}$ in bulk solution. As shown in Fig. 2a, compared with PMS alone system, negligible improvement of TC removal efficiency was attained by leached $Co^{2+}$, which excluded the possibility that homogeneous catalytic reaction dominated TC removal in $CoFe_2O_4$ or $CoFe_2O_4$@PC system. Moreover, the impact of different As(III) concentrations was investigated. Increasing As(III) concentration could be inhibitory for TC degradation. With 100 μM As(III), TC removal efficiency was decreased to 86.0% in the $CoFe_2O_4$@PC system. The results indicate that As(III) and TC may compete for the oxidation capacity of the $CoFe_2O_4$@PC system, though relatively high TC removal performance could still be achieved in the existence of As(III).

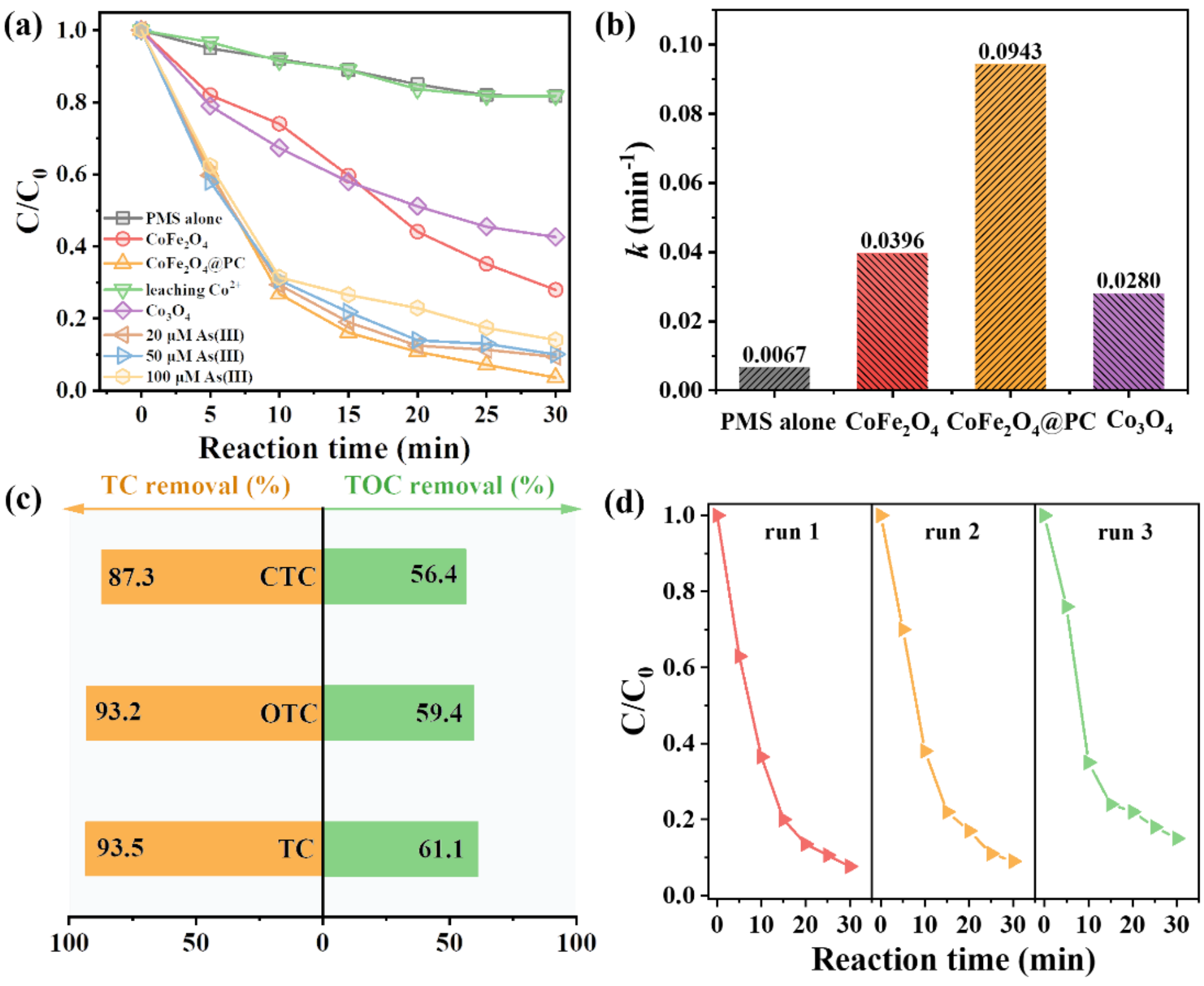

**Fig. 2.** TC removal in various PMS activation systems (a) and corresponding kinetic constant (b); Degradation and mineralization efficiency of TC, CTC, and OTC (c); cycle tests for $CoFe_2O_4$@PC (d). Experimental conditions: [TC] = [CTC] = [OTC] = 10 mg/L, [PMS] = 0.8 mM, [Catalysts] = 0.1 g/L, initial pH = 7.

The capability of $CoFe_2O_4$@PC in degrading other tetracycline antibiotics, including CTC and OTC, was explored. Fig. 2c shows that degradation efficiency could reach 93.2% and 87.3% for OTC and CTC within 30 min, respectively. Moreover, the TOC removal efficiency was 61.1% for TC, 59.4% for OTC, and 56.4% for CTC within 30 min. The results illustrated that $CoFe_2O_4$@PC showed an excellent catalytic performance in removing a series of tetracyclines. To evaluate the reusability and stability of $CoFe_2O_4$@PC, we recycled the catalyst with a membrane filter after each run of TC degradation, washed it with DI water to remove PMS and organic residues, and dried it at 80 °C in the oven. Degradation tests were repeated under the same condition in three successive runs. The catalytic system could still remove 85% of TC in the third run (Fig. 2d), and, due to the solid magnetic ability of $CoFe_2O_4$@PC, the catalysts could be easily recycled with a magnet (Fig. S2), indicating the excellent stability and reusability of $CoFe_2O_4$@PC in the Fenton-like degradation process. ICP-OES was used to examine the extent of metal leaching in the reacted solution. Results in Table S1 showed that the leaching amount of Co was 1.15 mg/L and that of Fe was undetectable in the $CoFe_2O_4$@PC system, which was much lower than the $CoFe_2O_4$ system (6.288 mg/L Co and 0.236 mg/L Fe). The result proved that the intense interaction between $CoFe_2O_4$ and PC drastically inhibited the leaching of Co and Fe. The TC degradation experiment demonstrated that $CoFe_2O_4$@PC possesses excellent catalytic performance and stability. The influence of distinct factors on catalytic activity was studied further to evaluate the potential of the $CoFe_2O_4$@PC system.

The effects of different impact factors on TC degradation kinetic rates were explored and presented in Fig. 3. The impact of the initial solution pH, within the range of 3.0-11.0, was investigated first. As the pH increased from 3.0 to 7.0, the TC removal rate rose from 0.051 to 0.094 $min^{-1}$. With solution under acidic conditions, $HSO_5^-$ tends to convert to $H_2SO_5$, inhibiting the production of $\cdot SO_4^-$ and ·OH radical species [31]. When the pH continued to increase to 9.0-11.0, a noticeable decrease in degradation efficiency could be observed because excessive $OH^-$ may scavenge $\cdot SO_4^-$ and ·OH, reducing the oxidation efficiency of the system [32,33]. Overall, contented degradation efficiency with different solution pH was attained, suggesting a wide pH range.

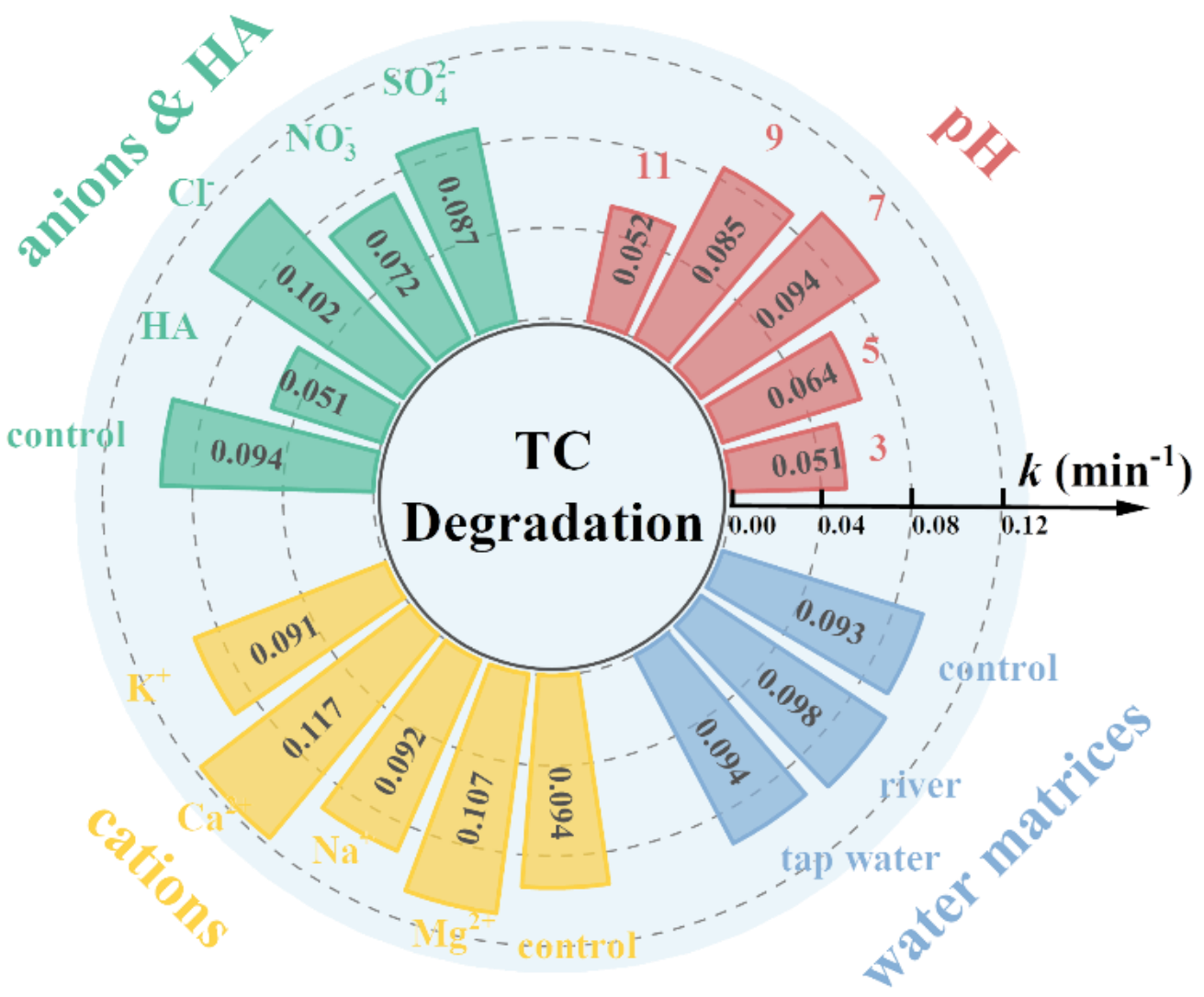

**Fig. 3.** Effects of pH, HA, anions, cations and water matrices on TC degradation. Reaction conditions: [TC] = 10 mg/L, [PMS] = 0.8 mM, [Catalysts] = 0.1 g/L, [anions] = [HA] = [cations] = 10 mM.

In aquatic environments, inorganic ions and natural organic matter could accelerate or inhibit Fenton-like oxidation. Catalytic performance was evaluated with $SO_4^{2-}$, $NO_3^-$, $Cl^-$ and humic acid (HA). Results revealed that anions had a slight impact on TC degradation. On the other hand, HA impeded TC degradation because the hydroxyl group of the HA molecules was favorably oxidized by the activated PMS, delaying TC degradation [34]. Moreover, the effect of cations ($Na^+$, $K^+$, $Ca^{2+}$, and $Mg^{2+}$) on TC degradation was also revealed. The monovalent ions ($K^+$ and $Na^+$) weakly affected TC removal efficiency. Notably, the divalent cations ($Ca^{2+}$ and $Mg^{2+}$) positively accelerate TC oxidation. Reports have confirmed that $Ca^{2+}$ and $Mg^{2+}$ have strong complexation with TCs, and the formation of metal-TC complexes enhanced the degradation of TC by PMS [35–37]. Furthermore, the influence of different water matrices was also explored. The tap water was collected from the college campus, and the river was drawn from the Pearl River in Guangzhou. Different water matrices showed negligible effect on TC degradation, indicating the system has good prospects for water treatment application.

*3.2.2. As(III) removal*

Removal of As(III) by $CoFe_2O_4$@PC was then studied. As shown in Fig. 4a, the $CoFe_2O_4$@PC system removed 91% of As(III) without TC, with only a 5% decrease in removal efficiency when TC was present. Similarly, 49% of As(III) was removed in the PMS alone system with the coexistence of TC, with only a 2% efficiency reduction compared to the PMS alone system without TC. The results suggest that the coexistence of TC slightly inhibited As(III) removal. Moreover, the oxidation of As(III) could be divided into two periods: rapid oxidation of As(III) within 1 min and slow oxidation in the continued process, which was consistent with previously reported work [38]. The result of that insignificant effect of TC on As(III) oxidation revealed that the reaction rate of As(III) oxidation was much higher than that of TC degradation in the first period.

To further elucidate the impact of TC on As(III) removal, As(III) oxidation was analyzed in combination, and the results confirmed the negligible effect of TC on As(III) oxidation. Differences between As(III) removal and oxidation efficiency were mainly due to the effective adsorption brought by the large surface area of porous carbon and the strong affinity of spinel ferrites for As species.

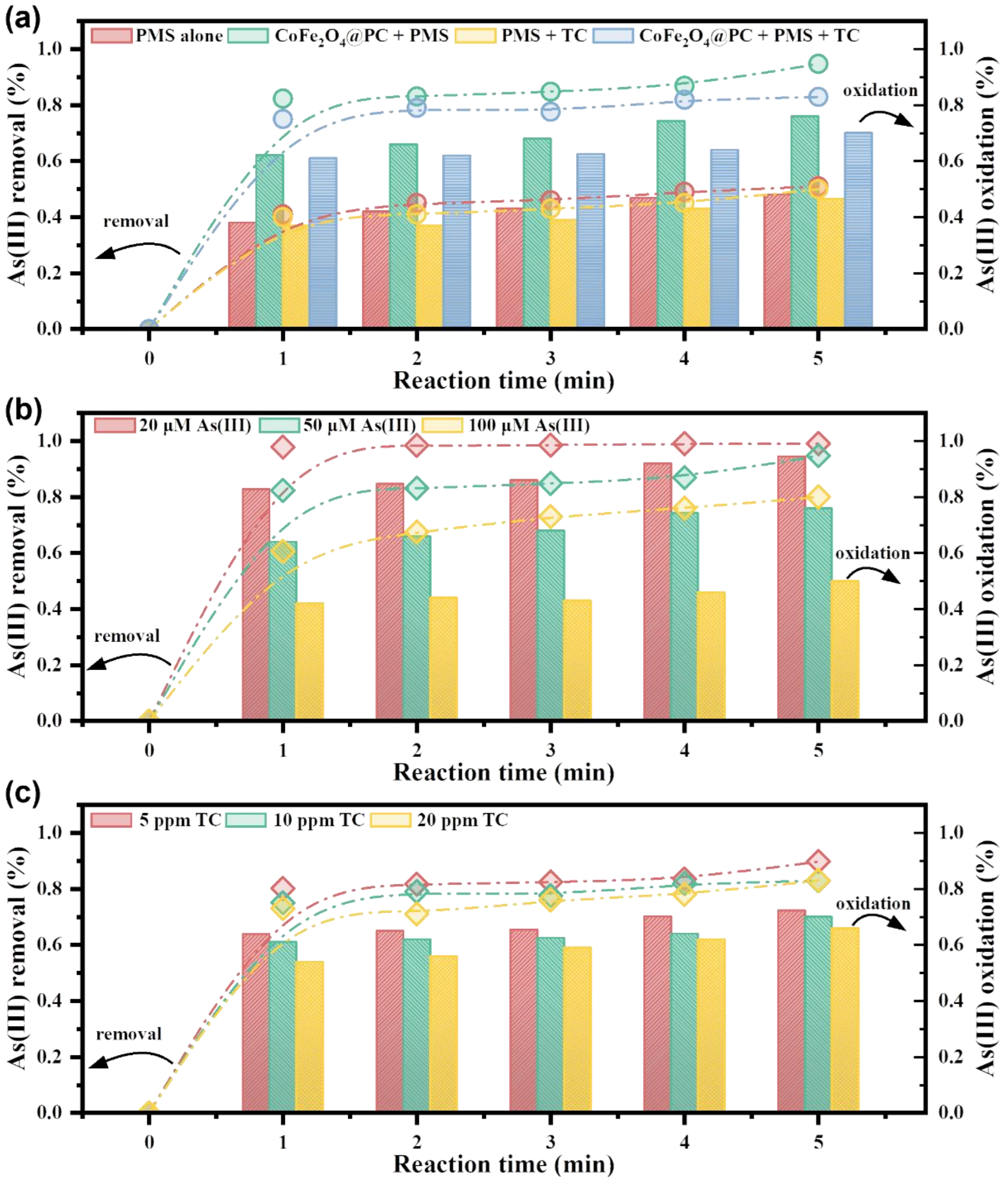

**Fig. 4.** As(III) removal efficiency and oxidation efficiency with different reaction systems ((a) and (d)), different As(III) concentrations ((b) and (e)), and different TC

concentrations ((c) and (f)). Quenching experiments for $CoFe_2O_4$@PC/PMS/As(III) system (c). Experimental conditions: [As(III)] = 20 μM, [PMS] = 0.8 mM, [Catalysts] = 0.1 g/L, initial pH = 7.

As shown in Fig. 4b, As(III) concentrations significantly affect As(III) removal. increasing As(III) concentration from 20 μM to 100 μM, As(III) removal efficiency decreased from 99% to 80%, and As(III) oxidation efficiency decreased from 94% to 50%. Moreover, the influence of different TC concentrations on As(III) removal was presented in Fig. 4c. Limited reduction in As(III) removal and oxidation efficiency revealed a negligible inhibiting effect on As(III) removal. The results suggest that As(III) concentration was the main limiting factor for As(III) removal efficiency.

### *3.3. Catalytic mechanisms*

To unveil the contribution of reactive oxygen species (ROS) to TC removal in the $CoFe_2O_4$@PC system, quenching experiments were performed with methanol (MeOH), tert-Butyl alcohol (TBA), and L-histidine. MeOH can scavenge ·OH ($9.7 \times 10^8$ $M^{-1}$ $s^{-1}$) and $\cdot SO_4^-$ ($2.5 \times 10^7$ $M^{-1}$ $s^{-1}$). TBA can react effectively with ·OH ($6 \times 10^8$ $M^{-1}$ $s^{-1}$), and L-histidine was used to scavenge $^1O_2$ ($3.2 \times 10^7$ $M^{-1}$ $s^{-1}$) [39]. As shown in Fig. 5a, the degradation performance of TC was inhibited by 40% with the addition of MeOH, while the addition of TBA decelerated TC degradation with only 60% removal efficiency. These results indicate that ·OH contributes significantly to TC degradation, while $\cdot SO_4^-$ plays a minor role. On the other hand, adding 10 mM L-histidine caused a 42% reduction in TC degradation efficiency, revealing the essential involvement of $^1O_2$. Increasing quenchers were added to compare TC removal efficiency in 30 min to ensure

ROS was fully quenched. As shown in Fig. 5b, the final removal efficiency of TC changed little as the quencher dosage increased, confirming the validity of the results in Fig. 5a. The results showed that ·OH and $^1O_2$ played a key role in TC degradation. At the same time, $\cdot SO_4^-$ contributed less than expected to the process. Besides, the experiments also implied the coexistence of radical and nonradical pathways, which can lead to higher degradation efficiency of TC than a single pathway [40].

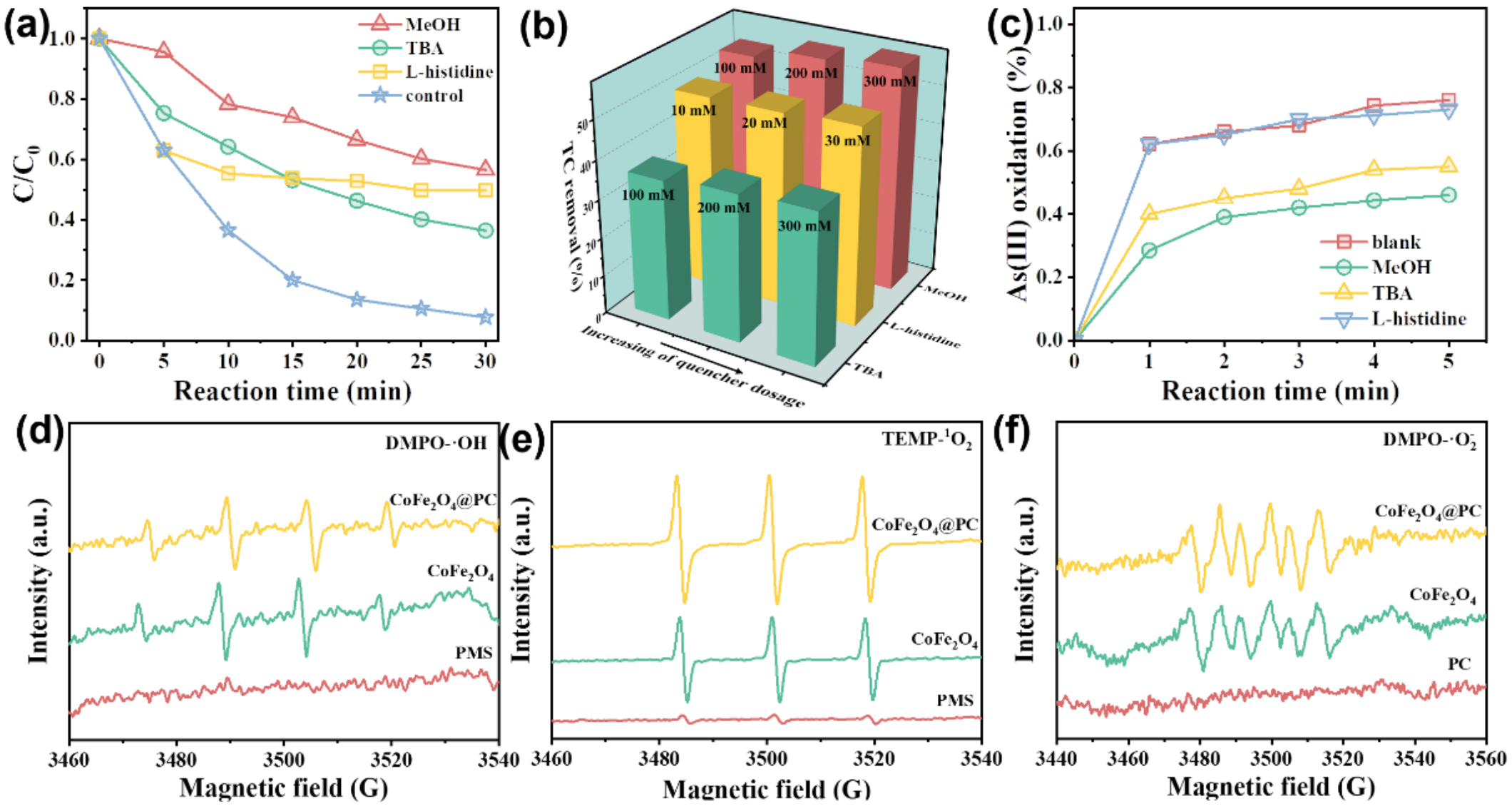

**Fig. 5.** The quenching experiments of TC degradation in the $CoFe_2O_4$@PC/PMS system (a) and the impact of increasing quencher dosage (b); The quenching experiments of As(III) oxidation in the $CoFe_2O_4$@PC/PMS (c). [MeOH] = [TBA] = 100 mM, 200 mM, 300 mM, [L-histidine] = 10 mM, 20 mM, 30 mM; ESR spectra in the presence of DMPO (d) and TEMP (e) in water, and DMPO in methanol (f).

Similarly, trapping experiments were also conducted to clarify the contribution of ROS to As(III) oxidation. As shown in Fig. 5c, it could be found that As(III) oxidation efficiency declined from 76% to 46% and 55% in the presence of MeOH and TBA, respectively. However, the inhibitory effect of 10 mM L-histidine was negligible, implying that $^1O_2$ merely contributed to the oxidation of As(III). Therefore, As(III) was oxidized via a radical pathway in the $CoFe_2O_4$@PC system.

Electron spin resonance (ESR) tests were conducted with DMPO and TEMP as spin-trapping reagents to validate ROS's roles in the system further. As shown in Fig. 5d, the quartet signals of DMPO-·OH were identified both in the $CoFe_2O_4$@PC and $CoFe_2O_4$ system with similar intensity, which verified the existence of ·OH in both system, while negligible signals were observed in the PMS alone system. The results confirmed that $CoFe_2O_4$ was the main active sites for radical production. On the other hand, DMPO-·$SO_4^-$ were merely seen, suggesting the minor role of ·$SO_4^-$ in the degradation process. Besides, Fig. 5e exhibited the typical triplet signals of TEMP-$^1O_2$ in the PMS alone, $CoFe_2O_4$ and $CoFe_2O_4$@PC system. Specifically, in the PMS alone system, a weak trace of $^1O_2$ was due to the self-decomposition of PMS [39]. Notably, $^1O_2$ signals in the $CoFe_2O_4$@PC system were more intense than that in $CoFe_2O_4$ system, which suggests that, as reported before, PC may accelerate electron transfer between $CoFe_2O_4$ and PMS by increasing the conductivity of the catalyst [41–43]. Moreover, it was reported that $^1O_2$ could be generated via a reaction between ·$SO_5^-$ [44] or the coupling of superoxide radicals (·$O_2^-$) and ·OH [45]. Therefore, to confirm the existence of ·$O_2^-$ in the $CoFe_2O_4$@PC system, DMPO was used as capturing agent in methanol solution, and the results in Fig. 5f confirm the weak intensity of DMPO-·$O_2^-$ signals, indicating part of $^1O_2$ was generated via reaction between ·$O_2^-$ and ·OH. The above results revealed that the radical pathway (·OH, ·$SO_4^-$, ·$O_2^-$) and the nonradical pathway ($^1O_2$) both play essential roles in TC and As(III) oxidation. The cooperative effect enhanced oxidation efficiency and benefited the mineralization of organic pollutants, which would be mentioned afterward with intermediates analysis.

To investigate the synergetic effect between Co and Fe in $CoFe_2O_4$@PC, X-ray photoelectron spectroscopy (XPS) was used to examine the surface element valence change of Co and Fe. As shown in Co 2p (Fig. 6a), the three peaks positioned at 779.0 eV, 780.2 eV, and 781.9 eV can be attributed to octahedral $Co^{2+}$ ($Co^{2+}_{Oct}$), tetrahedral $Co^{2+}$ ($Co^{2+}_{Tet}$) and octahedral $Co^{3+}_{Oct}$, respectively [46,47]. For the fresh catalyst, the relative contributions of $Co^{2+}$ and $Co^{3+}$ to the total Co intensity were 67.2% and 32.8%, respectively. High $Co^{2+}/Co^{3+}$ ratio implied more $Co^{2+}$ active sites for efficient PMS activation. For the spent catalyst, the relative contributions of $Co^{2+}$ and $Co^{3+}$ were 61.6% and 38.4%, respectively. About 5.6% of $Co^{2+}$ converted into $Co^{3+}$, indicating a redox cycle of $Co^{2+}$-$Co^{3+}$-$Co^{2+}$ may occur. Notably, the content of $Co^{2+}_{Oct}$ decreased more significantly than that of $Co^{2+}_{Tet}$. This variation indicates that $Co^{2+}_{Oct}$ played an essential role in the activation of PMS. Generally, The overlap between O 2p orbitals of PMS and Co 3d orbitals has an essential impact on PMS activation [48]. $Co^{3+}_{Oct}$ ($t_{2g}^6e_g^0$) has no unpair electron, while $Co^{2+}_{Tet}$ ($e^4t_2^3$) and $Co^{2+}_{Oct}$ ($t_{2g}^6e_g^1$) both possess unpair electrons in 3d orbitals, making it the main active centers for PMS activation. For $Co^{2+}_{Tet}$, its unpaired electrons are in $t_2$ orbitals ($d_{xy}$, $d_{yz}$, $d_{zx}$), which cannot overlap with O 2p orbitals spatially or only form weak π bonds. In contrast, for $Co^{2+}_{Oct}$, its unpaired electrons are in $e_g$ orbitals ($d_{z2}$, $d_{x2-y2}$), which can form head-on overlap with O 2p [49,50], reinforcing the interaction between PMS and $Co^{2+}_{Oct}$. Therefore, $Co^{2+}_{Oct}$ is more active than $Co^{2+}_{Tet}$ in the Fenton-like process.

For Fe 2p spectra (Fig. 6b), the two peaks located at 710.2 eV and 712.7 eV can be attributed to $Fe^{3+}$ and $Fe^{2+}$, respectively [47]. The fresh sample contained 49.4% of $Fe^{3+}$ and 50.6% of $Fe^{2+}$. For the spent sample, the content of $Fe^{2+}$ decreased to 46.1%, indicating the participation of Fe in PMS activation. With relatively low redox potential ($Fe^{3+}/Fe^{2+}$, 0.771 V), the activation of PMS ($HSO_5^-/\cdot SO_4^-$, 1.82 V) with $Fe^{2+}$ is thermodynamically favorable [51,52]. Besides, the redox cycle between Fe and Co is also a thermodynamically feasible process for facilitating PMS activation, which also benefits the stability of $CoFe_2O_4$.

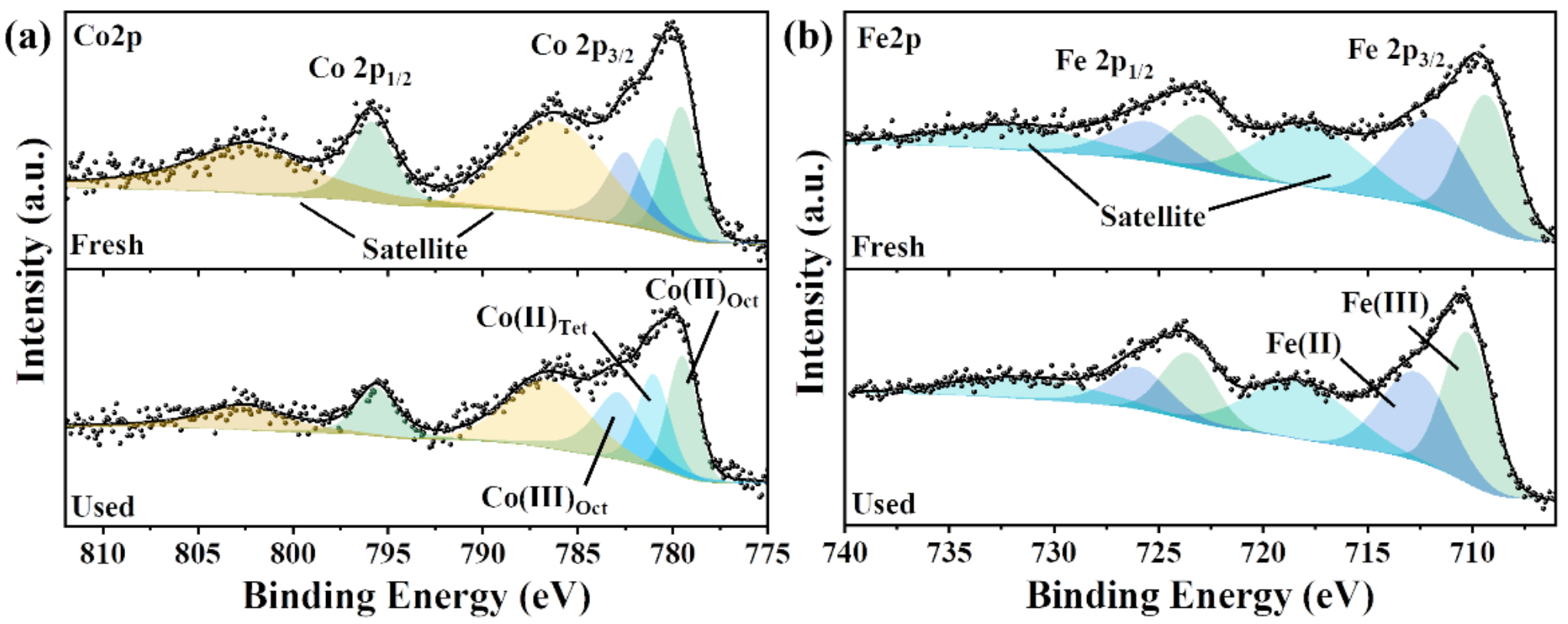

**Fig. 6.** Co 2*p* (a) and Fe 2*p* (b) XPS spectra of the fresh and used $CoFe_2O_4$@PC.

Based on the above results, a plausible mechanism for the simultaneous removal of As(III) and TC was proposed. Firstly, TC and As(III) were efficiently adsorbed to the surface of $CoFe_2O_4$@PC owing to the porous structure and large specific surface area of porous carbon. $Co^{2+}$ and $Fe^{2+}$ could reduce PMS to produce $\cdot SO_4^-$ (Eqs. 1-2). More radicals, including ·OH and $\cdot O_2^-$, were generated with chain reactions (Eqs. 3-4), then $Co^{2+}$ and $Fe^{2+}$ were concurrently converted to $Co^{3+}$ and $Fe^{3+}$, respectively. On the other hand, $Fe^{3+}$ and $Co^{3+}$ could also oxidize PMS and form $\cdot O_2^-$ and $\cdot SO_5^-$ (Eqs. 5-7), which then converted to $^1O_2$ (Eq. 8-9). Besides, the redox cycle of Co and Fe facilitated the regeneration of $Co^{2+}$ and $Fe^{2+}$ (Eq. 10), thus increasing the PMS activation rate and pollutant removal. These ROS cooperatively oxidized As(III) and TC, and As(III) was converted to As(V) (Eqs. 11-12) and adsorbed on catalysts. With the solid magnetic property of $CoFe_2O_4$@PC, As(III) and As(V) could be easily recycled and removed entirely from water. Simultaneously, TC was oxidized and mineralized to small products such as $H_2O$ and $CO_2$.

$$Co^{2+} + HSO_5^- \rightarrow Co^{3+} + \cdot SO_4^- + OH^- \quad (1)$$

$$Fe^{2+} + HSO_5^- \rightarrow Fe^{3+} + \cdot SO_4^- + OH^- \quad (2)$$

$$\cdot SO_4^- + H_2O \rightarrow \cdot OH + SO_4^{2-} + H^+ \quad (3)$$

$$\cdot SO_4^- + OH^- \rightarrow \cdot OH + SO_4^{2-} \quad (4)$$

$$Fe^{3+} + HSO_5^{2-} + H_2O \rightarrow Fe^{2+} + \cdot O_2^- + SO_4^{2-} + H^+ \quad (5)$$

$$Co^{3+} + HSO_5^- \rightarrow Co^{2+} + \cdot SO_5^- + H^+ \quad (6)$$

$$Fe^{3+} + HSO_5^- \rightarrow Fe^{2+} + \cdot SO_5^- + H^+ \quad (7)$$

$$\cdot O_2^- + \cdot OH \rightarrow {}^1O_2 + OH^- \quad (8)$$

$$\cdot SO_5^- + \cdot SO_5^- \rightarrow {}^1O_2 + \cdot SO_4^- \quad (9)$$

$$Fe^{2+} + Co^{3+} \rightarrow Fe^{3+} + Co^{2+} \quad (10)$$

$$As(III) + \cdot SO_4^- \rightarrow As(IV) + SO_4^{2-} \quad (11)$$

$$As(IV) + \cdot SO_4^- \rightarrow As(V) + SO_4^{2-} \quad (12)$$

*3.4. TC degradation intermediates and toxicity evaluation*

To investigate TC degradation pathways, liquid chromatography–mass spectrometry (LC-MS) was employed to identify possible intermediates products of TC degradation. Possible intermediates were presented in Fig. S3 and Table S2. Previous reports have calculated the vulnerable sites on the TC molecule and their tendency to be attacked [40]. Generally, as shown in Fig. S4, C1, C9, and C16 are favorable for nonradical attacks triggered by $^1O_2$. C12 and C14 tend to be attacked by radicals, including $\cdot SO_4^-$ and $\cdot OH$, and the N-methyl group on C7 can be attacked in both pathways. Therefore, possible degradation pathways were summarized in Fig. S4, showing how radical and nonradical routes cooperatively degraded TC. Firstly, P1 (TC, m/z 445) could be attacked by $\cdot OH$ and $\cdot SO_4^-$ and converted to P2 (m/z 460) with the addition of hydroxyl on C13 and oxidation of hydroxyl on C12. Besides, P1 could also be attacked by $^1O_2$ and converted to P3 (m/z 476) via the addition of hydroxyl on C1 and C9, coupling with the oxidation of hydroxyl on C8. P4 (m/z 387) and P5 (m/z 416) were obtained via coupling of radical and nonradical pathways with the addition of hydroxyl on C9 and C14 or loss of N-methyl group on C7. For further degradation of intermediates, $\cdot SO_4^-$, $\cdot OH$, and $^1O_2$, could attack different sites on TC, thus improving the degree of mineralization. For example, P6 (m/z 413), P10 (m/z 384), and P11 (340) were obtained through multiple steps of oxidation and ring-open reaction. The double bond at the C1-C2 position of P2 could be oxidized by $^1O_2$, producing a ketone group and a carboxylic group, and further oxidation via both pathways destructed the benzene structure of P2, which turned into P6 [53]. Further oxidation of P6-P12 converted intermediates into smaller molecules, including P13-P17, and all byproducts eventually decompose into smaller molecules and finally convert to $CO_2$ and $H_2O$.

Toxicity Estimation Software Tool (TEST) was utilized to evaluate the toxicity of TC and degradation intermediates. The $LC_{50}$ of fathead minnow and daphnia magna were two important indicators for the acute toxicity. As shown in Fig. S5a-b, TC was very toxic to fathead minnow and toxic to daphnia magna. However, $LC_{50}$ of most degradation intermediates were much lower than that of TC, suggesting the remarkable reduction of toxicity impact on environment. Although in Fig. S5c, TC and most intermediates were still identified as toxicants for biological development, the developmental toxicity of most intermediates showed varying degrees of reduction. These results suggest that the overall toxicity was effectively reduced in the $CoFe_2O_4$@PC system. Overall, this work provides a new strategy for the simultaneous removal of TC and As(III) in livestock wastewater (Fig. 7).

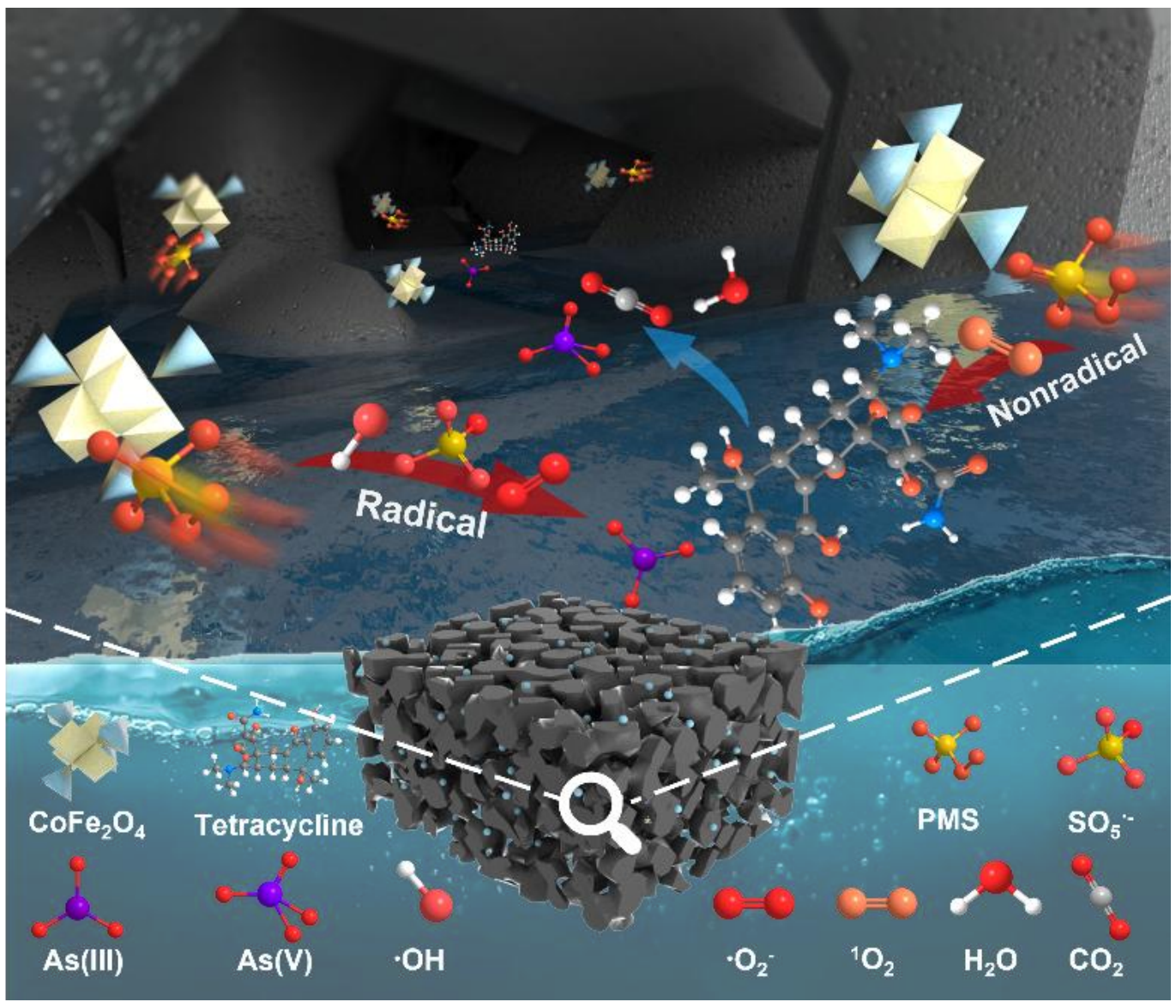

**Fig. 7.** Simultaneous As(III) and TC removal in the $CoFe_2O_4$@PC/PMS system.

## 4. Conclusions

In this work, the novel magnetic $CoFe_2O_4$@PC was facilely prepared and anchored on biomass-derived porous carbon via a facile microwave-assisted calcination method. $CoFe_2O_4$ nanoparticles were finely anchored on porous carbon, which formed strong interaction between $CoFe_2O_4$ and porous carbon, and more active sites were exposed than that of pure $CoFe_2O_4$. The optimized $CoFe_2O_4$@PC system could simultaneously remove 96% of TC (10 mg/L) and 86% of As(III) (50 μM) within 30 min, respectively. More importantly, 70% of As(III) could be oxidized into nontoxic As(V). Though the coexistence of TC and As(III) competed for removal capacity, the $CoFe_2O_4$@PC system still exhibited superior performance with efficient adsorption and oxidation activity due to the following features: (i) Fast redox cycle of Co and Fe in $CoFe_2O_4$@PC was beneficial for the regeneration of Co(II) and Fe(II); (ii) Close interaction between $CoFe_2O_4$ and PC facilitate electron transfer between $CoFe_2O_4$ and PMS; (iii) The PC could boost the mass transfer efficiency and protect $CoFe_2O_4$ against Co and Fe ions leaching. Reactive oxygen species, including $\cdot OH$, $\cdot SO_4^-$, $\cdot O_2^-$ and $^1O_2$, were detected in the $CoFe_2O_4$@PC system, confirming the radical and nonradical pathway cooperation in TC degradation, while the quenching experiment showed that As(III) was majorly oxidized in the radical pathway. Besides, degradation intermediates of TC were identified and proposed degradation pathways illustrated how radical and nonradical species reacted with TC. Overall, the remarkable activity and stability of $CoFe_2O_4$@PC offer immense potential for treating antibiotics and heavy metals in livestock wastewater.

**Acknowledgement**

This work was supported by National Natural Science Foundation of China (22078374), the National Ten Thousand Talent Plan, Key Realm Research and Development Program of Guangdong Province (2020B0202080001), Science and Technology Planning Project of Guangdong Province, China (2021B1212040008), the Scientific and Technological Planning Project of Guangzhou (202206010145).